\documentstyle[12pt]{article}

\begin{document}
\begin{titlepage}
\begin{center}
\vspace*{1.5cm}
\hfill
\vbox{
    \halign{#\hfil         \cr
           28 October 1997\cr} 
          }
\vskip 1cm
{\large \bf THE RAYLEIGH-TAYLOR INSTABILITY OF TRANSITION LAYER }
\vskip 1cm
A.Tavakoli\footnote{e-mail:tavakoli@theory.ipm.ac.ir}$^{\dagger}$,
D.D.Tskhakaya$^{\ddagger}$

$^{\dagger}${\small\it Institute for Studies in Theoretical Physics and Mathematics P.O.Box
19395-5531,Tehran,Iran.}  \\
$^{\ddagger}${\small\it Institute of Physics,Georgian Academy of Sciences P.O.Box 380077,
Tbilisi,Georgia.}
\end{center}
\vskip 1cm

\abstract
{New types of symmetry for the Rayleigh equation are found. For small
Atwood number, an analytic solution is obtained for a smoothly varying density
profile. It is shown that a transition layer with a finite width can
undergo some kind of stratification.}
\end{titlepage}
\newpage
\par Although Rayleigh had investigated the instability (known now as
Rayleigh-Taylor (RT) instability) using an exponential profile of density
\cite{1}, most investigations have been done
so using sharp interfaces [2-6].
In \cite{7} it is carried out making an interesting analogy between the
equations which describes the RT instability, and the Schrodinger equation.
In that study, an \lq\lq equivalent potential" is constructed and for
well-known density profiles it is shown that the Schrodinger equation with
the corresponding \lq\lq equivalent potential" has the same eigenvalues as
the Rayleigh equation.
\par The physical quantities (e.g.; density, velocity) and their derivatives,
generally speaking cannot suffer a jump discontinuity. Therefore, it seems
more consistent to attempt to solve the problem for a transition layer of
finite thickness and then take the limit when the thickness of this layer
$\Delta$ tends to zero; i.e. consider the case of a density jump in the
limit, $\Delta \rightarrow 0$. The solution for this transition layer gives
us a more complete physical picture of the instability.
\par In this paper we consider the RT instability of a transition layer of
finite thickness, where the unperturbed density changes continuously from a
constant value up to another one. For that, an analytic solution is found, in
the limit of small Atwood number. This solution allows us to investigate the
$\Delta\rightarrow 0$ limit. We find that:
\par 1. For an arbitrary density profile the \lq\lq equivalent potential" can be
presented  in compact form. This allowed us to predict a new type of
symmetry (in addition to the well-known symmetry of the Rayleigh equation
\cite{2}).
\par 2. The depth of the \lq\lq equivalent potential" well depends on the width
of the transition layer and the depth not the {\it energy} of the potential
well undergoes a quantization. For the finite width of the layer the
spectrum of eigenvalues is infinite. This conclusion is consistent with the
fact that with the increase of the quantum number grows the well depth.
\par 3. The eigenfunctions found here show the exfoliation of the transition
layer.

\par I. Using the linearized equation of motion and the continuity equation
with the help of incompressibility condition leads one to:
\begin{equation}
      {\rho_0(z)\frac{\partial^2 v(z,t)}{\partial t^2}={1\over k^2}\frac
      {\partial^2}{\partial t^2}\frac{\partial}{\partial z}[\rho_0(z)
      {\frac{\partial v(z,t)}{\partial z}}]+
      gv(z,t)\frac{\partial \rho_0(z)}{\partial z}}   ,
\end{equation}
where $\rho_0(z)$ describes the equilibrium density profile, $v(z,t)$ is the
$z$ component of the perturbed fluid velocity, $k$ is the wave number
of the perturbed quantities which are chosen proportional to $e^{iky}$
and $g$ is the constant acceleration along negative $z$. Then one
can find the solution of the above equation using separation of variables:
\begin{equation}
       v(z,t)=T(t)\Phi(z).
\end{equation}
Then Eq. (1) reduces into two equations as follows:
\begin{eqnarray}
          T^{\prime\prime}(t)-\gamma^2T(t)=0   ,    \\
   {\Phi^{\prime\prime}(z)+{\rho_0^\prime(z)\over \rho_0(z)}\Phi^\prime(z)
   +(-k^2+{k^2g\over \gamma^2}{\rho_0^\prime(z)\over \rho_0(z)})\Phi(z)=0} ,
\end{eqnarray}
where $\gamma^2$ is the constant of separation, primes denote derivatives
with respect to corresponding arguments. The case $\gamma^2 > 0$
corresponds to unstable modes.
\par The substitution
\begin{equation}
       \Phi(z)={\psi(z)\over \sqrt{\rho_0(z)}}
\end{equation}
reduces Eq. (4) to the form of the Schrodinger equation as in \cite{7}:
\begin{equation}
      \psi^{\prime\prime}(z)+[-k^2-V(z)]\psi(z)=0    ,
\end{equation}
with the \lq\lq equivalent potential",
\begin{equation}
     {V(z)={1\over \sqrt{\rho_0(z)}}\{\frac{\partial^2}{\partial z^2}-2k\Gamma
     \frac{\partial}{\partial z}\}\sqrt{\rho_0(z)}}    ,
\end{equation}
where
\begin{equation}
       \Gamma={kg\over \gamma^2}  .
\end{equation}
One can further simplify Eqs. (6) , (7) as follows:
\begin{eqnarray}
    V(z)={S^{\prime\prime}(z)\over S(z)}-k^2\Gamma^2,        \\
   {{\psi^{\prime\prime}(z)\over \psi(z)}-k^2={S^{\prime\prime}(z)\over
   S(z)}-k^2\Gamma^2,}
\end{eqnarray}
where
\begin{equation}
    S(z)=e^{-k\Gamma z} \sqrt{\rho_0}.
\end{equation}
This form then allows us to prove the following symmetry properties of the
Rayleigh equation for arbitrary density profiles:\par
a) The substitutions
\begin{equation}
       {\Gamma \rightarrow -\Gamma\ \ \ {\rm and}\ \ \ \rho_0(z) \rightarrow
       \rho_0(z)e^{-4k\Gamma z}}
\end{equation}
cause Eq. (9), and consequently Eq. (6), to remain unchanged. This means
that every unstable mode, $\Gamma > 0$, with the density profile $\rho_0(z)$
can be compared with a stable, oscillating mode, $\Gamma < 0$ (see Eqs. (3)
, (8)), with the density profile $\rho_0(z)e^{-4k\Gamma z}$ and vice versa.
For the modified profile, on the other hand, the expression for the velocity
(see Eqs. (2) , (5)) will contain an additional factor $e^{2k\Gamma z}$.
\par b) Performing the substitution $\rho_0(z)$ by $\rho_0(-z)e^{4k\Gamma z}$
and then the transformation $z \rightarrow -z$, we find that the potential (9)
and Eq. (6) remain unchanged. Consequently the spectrum of the eigenvalues
for both profiles are the same, whereas eigenfunction of the latter can be
obtained by sending $z\rightarrow -z$ in the former. Furthermore, the function
$\Phi(z)$ upon transformation becomes $e^{-2k\Gamma z}\Phi(z)$. Note that
these symmetry properties hold only for fixed $\Gamma_n (n=0,1,2,...)$

\par II. Let us assume that the continuous
equilibrium density tends to different constant limits, i.e. $\rho_0(z)
\rightarrow \rho_1$ at $z \rightarrow -\infty\ \ $ and $\ \ \rho_0(z)
\rightarrow \rho_2$ at $z \rightarrow +\infty$, where $\ \ \rho_2 > \rho_1\
\ $, and $\rho_0^\prime(\pm\infty)=0$. Then if somewhere in the region
$-\infty<z<+\infty$ the \lq\lq equivalent potential" becomes negative,
according to Eq. (6) the RT problem becomes analogous to a particle's motion
in a potential well. Using a tanh function representation for density:
\begin{equation}
       \rho_0(z)=c(1+ A\cdot \tanh{{2z\over \Delta}})    ,
\end{equation}
where
\begin{equation}
     c={\rho_2+\rho_1\over 2}\ \ ,\ \ A={\rho_2-\rho_1\over\rho_2+\rho_1},
\end{equation}
enable us to examine different limits, from an smooth to a sharp jump limit.
We restrict ourselves to the case of small Atwood number ,
$A\ll 1$, where one can obtain an analytic solution. Neglecting the terms of
order $A^2$ from Eq. (6) we obtain:
\begin{equation}
     {\psi^{\prime\prime}(x)+\{-\delta^2+{1\over \cosh^2{x}}[\delta\cdot A\Gamma
     +A\cdot \tanh{x}]\}\psi(x)=0,}
\end{equation}
where $\delta=k\Delta/2\ \ $,$\ \ x=2z/\Delta$.
We made an assumption that $A\Gamma \geq 1$, which will be confirmed by
the results. For a broad transition layer, i.e. $\delta \gg A$, Eq. (15)
can be reduced to a form which is well-known in quantum mechanics \cite{8}
(see p. 73):
\begin{equation}
     {\psi^{\prime\prime}(x)+[-\delta^2+{\delta\cdot A\Gamma\over \cosh^2{x}}]
     \psi(x)=0}
\end{equation}
Solution to this equation finite for $z \rightarrow\pm\infty$, can be
expressed in terms of hypergeometric function $F$:
\begin{equation}
   {\psi = \cosh^{-\delta}{x}\ F(\delta-s,\delta+s+1,\delta+1,(1-\tanh{x})/2),}
\end{equation}
the parameters of which ($ \delta\ \  ,\ \ s={1\over 2}[-1+
\sqrt{1+4\delta A\Gamma}]$ ) have to satisfy the relation:
\begin{equation}
      \delta-s=-n,
\end{equation}
where $n=0,1,2,...$ . From Eq. (18) we find the spectrum of growth rates for RT
instability:
\begin{equation}
   {\gamma^2\over kg}={\delta\over (\delta+n)(\delta+n+1)}A.
\end{equation}
Consequently for a transition layer of finite width, $\delta \approx 1$,
there is an infinite spectrum of eigenvalues. Number of the extremum
and zero points of the eigenfunctions (17) is equal to $n+1$ and $n$ ,
respectively. For $n>0$ the spatial dependence of the eigenfunction has
oscillatory character. It means that due to the instability,
the monotonous distribution of the density in the transition layer can be
destroyed, and a stratified and multi-stream structure can be formed.
The velocity changes the direction as one goes from one new-formed
sublayer to another. The Figure shows the spatial distribution of the velocity
(determined from the function $\Phi (z)$; see Eq. (2)) and the total density $\rho(z)=\rho_0(z)+
\tilde{\rho}(z)$ in such a stratified layer. The density perturbation
$\tilde{\rho}(z)$ is obtained from the continuity equation:
\begin{equation}
   {\tilde{\rho}(z)=-\Phi(z)\frac{\partial \rho_0(z)}{\partial z}
   {1\over \gamma}\int_{0}^{\gamma t}d\tau T(\tau).}
\end{equation}
As we are concerned in the spatial dependence only, by plotting the figure we
treated the definite integral as constant (for simplicity we choose it as
equal to one). If the width of the transition layer is small, say
\begin{equation}
     \delta=k\Delta/2 \leq A \ll 1,
\end{equation}
we can consider such a layer as a jump in density. However, in this case the
analytic solution of Eq. (15) fails. It is necessary to note that although the
spectrum represented by Eq. (19) is obtained for a broad transition layer,
$\delta \gg A$, (when we neglect the second term in the square brackets in Eq.
(15)) Eq. (19) gives the correct expression for the main unstable mode
in the case of density jump, when $\delta\ll A$. In fact as $\delta
\rightarrow 0 $ from Eq. (19) it follows that the most unstable mode
corresponds to the quantum number $n=0$, and for the growth rate one obtains
the well-known expression \cite{1,2}:
\begin{equation}
    \gamma= \sqrt{kgA}.
\end{equation}
Such a general meaning of Eq. (19) can be explained by the fact that,
for $\delta \ll 1$ , $A \ll 1$ the \lq\lq equivalent potential" in Eq. (15)
can be considered as a perturbation. The quantum mechanical problem of
particle motion in a one-dimensional well, whose depth is small, is solved
in \cite{8} (see p. 162). There an equation for the determination of the
eigenvalues of the Schrodinger equation is obtained. That equation in our
notation becomes:
\begin{equation}
     -2\delta=\int_{-\infty}^{+\infty}dx V(x) .
\end{equation}
It is obvious that the second term of Eq. (15) doesn't give any contribution
in this integral (this term is odd). The first term gives the correct
expression for the growth rate consistent with Eq. (22). That is why for the
small \lq\lq equivalent potential", ($\delta \ll 1, A \ll 1$) the second term in square brackets in
Eq. (15) does not play any role in the determination of the growth rate and
the expression (19) is valid for a shallow well also.

\par We can summarize our results as follows:
\par 1. The new symmetry properties found in this paper show new density
profiles for which the problem of instability can be reduced to the one and
the same equation \cite{3,7,9}.
\par 2. According to the symmetry theorem known up to now, the two profiles
$\rho_0(z)$ and $1/\rho_0(-z)$ have the same set of eigenvalues but not
eigenfunctions \cite{2,3,9}. The symmetry properties found here maintain the
spectrum of eigenvalues and allow one to determine a relationship between
the eigenfunctions (see Eq. (5)).
\par 3. The Eq. (10) shows that the inverse problem (obtaining the density
profile for known perturbed quantity) is described by the same type of
equation as the original problem.
\par 4. From Eq. (17) it follows that for every eigenmode, there exists a
unique transition layer of stratified structure.
\par 5. It should be interesting to extend this analysis when separation of
variables is not used and thus the spatial and temporal parts not treated
independently.
\par These results are true not only for the chosen density profile from Eq.
(13) but also for any density profile whose corresponding
\lq\lq equivalent potential" has the form of a well.
\par We are indebted to prof. N.L.Tsintsadze for encouragement and
discussions. We thank Dr. F. Kazeminezhad for help in preparing of manuscript.

\end{document}